\documentclass[pre,twocolumn,showpacs,superscriptaddress]{revtex4-1}
\usepackage{amsfonts,amsmath,amssymb,graphicx}
\usepackage{amsfonts}
\usepackage{amsmath}
\usepackage{amssymb}
\usepackage{graphicx}
\newcommand{\f}{\frac}
\newcommand{\T}{\mathrm{t}}
\begin{document}

\title{\textbf{ Fidelity of the quantum delta-kicked accelerator} }
\author{R. K.  Shrestha}
\affiliation{Department of Physics, Oklahoma State University,
Stillwater, Oklahoma 74078-3072, USA}
\author{S. Wimberger}
\affiliation{Institut f\"{u}r Theoretische Physik, Universit\"{a}t Heidelberg, Philosophenweg 19, 69120 Heidelberg, Germany}
\author{J. Ni}
\affiliation{Department of Physics, Oklahoma State University, Stillwater, Oklahoma 74078-3072, USA}
\author{W. K.  Lam}
\affiliation{Department of Physics, Oklahoma State University, Stillwater, Oklahoma 74078-3072, USA}
\author{G. S. Summy}
\affiliation{Department of Physics, Oklahoma State University, Stillwater, Oklahoma 74078-3072, USA}

\begin{abstract}
\noindent The sensitivity of the fidelity in the kicked rotor  to an acceleration is experimentally and theoretically investigated. We used a Bose-Einstein condensate exposed to a sequence of  pulses from a standing light wave followed by a single reversal  pulse in which the standing wave was shifted by half a wavelength. The features of the fidelity ``spectrum" as a function of acceleration are presented. This work may find applications in the  measurement of  temperature of an ultra-cold atomic sample.
\end{abstract}

\maketitle
The study of non-linear systems is important to many branches of science. Consequently the chaotic behavior that they can exhibit in the classical regime has been extensively studied and used \cite{chaos,classicalchaos,quantumchaos}. A particularly interesting aspect of such systems is that due to the linearity of the Schr$\ddot{\text{o}}$dinger equation, their quantum and classical dynamics can be dramatically different. For this reason the so called delta-kicked rotor and it's quantum analog the quantum delta-kicked rotor (QDKR) have received much attention. The latter can be experimentally realized by subjecting a sample of cold atoms  to short pulses of an off-resonant standing wave of laser light  \cite{aoqr}.  The QDKR has  proved to be a paradigmatic model to study several important phenomena including quantum resonances (QR) \cite{resonance,resonance1},
dynamical localization \cite{resonance,localization}, and  quantum ratchets \cite{Ratchet}. A closely related system, the quantum delta kicked accelerator (QDKA), differs from the usual QDKR
by adding a linear potential in the form of an acceleration. The QDKA has been
 used in studying aspects of the transition to chaos in both classical and quantum regimes \cite{transition}, and is a system in which quantum accelerator modes \cite{fgr,gazal} are observed.

One of the common themes in the experiments mentioned above is that the quantum evolution is typically measured indirectly through observations of the momentum distribution. However recently it has become possible to study the coherent evolution  of a superposition
 of state vectors   directly  by examining the  overlap of the atomic state with a reference state. This quantity is termed ``fidelity". It  has garnered considerable interest as
an alternative way of studying   coherent evolution  in
the context of quantum-classical correspondence \cite{correspondence} and  quantum information processing \cite{information}.  Recently, it    was shown that the width of a pulse-period fidelity resonance of the QDKR exhibits  sub-Fourier scaling  \cite{njpfidelity,fidelity},    where the width of the resonance scales as the inverse cube of the number of applied pulses.
Because of this sensitivity to the pulse period, the fidelity technique was proposed  as a means for improving
the precision of frequency measurements \cite{fidelity}. Although  subsequent work has shown possible limitations with this approach \cite{sackett}, we show here that the observed asymmetry in the fidelity with respect to an acceleration may be used for temperature measurements of the atomic cloud.
\begin{figure}[h]
\includegraphics[width=9.0 cm, height=8.0 cm]{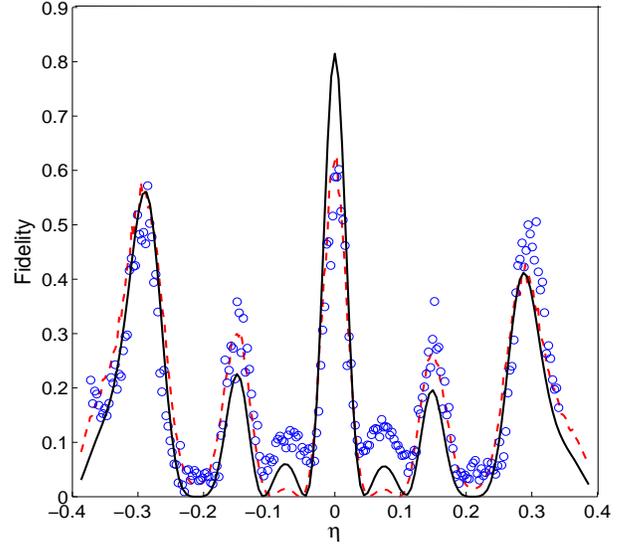}
\caption{(Color Online) Fidelity as a function of the scaled acceleration, $\eta$,  due to four kicks of strength $\phi_d\approx 0.6
$ followed by a reversal kick of strength $\approx 4\phi_d$. The  black solid (red dashed) line is a numerical simulation   with $\tau=2\pi$ (i.e. $\ell=1$), $\beta=0.5$ and initial momentum width $\Delta\beta=0.06\hbar G$ without (with) effects such as vibrations and reversal phase imperfections (see more in the text). Circles are experimental data. Note that the fidelity has a rich structure with multiple resonant peaks. All fidelity measurements are $\pm0.01$.}\label{longscan}
\end{figure}

In this paper we discuss  the sensitivity of fidelity in the QDKA to an externally applied acceleration. A full analytical theory (neglecting atomic interactions) along with corresponding experimental results and  numerical simulations
 are presented. We show that the width of resonant peaks in fidelity as a function of acceleration are sensitive to the momentum width of the atomic sample, the pulse period, and the direction of the acceleration.
\begin{figure}[h]
\includegraphics[width=8.0 cm, height=7cm]{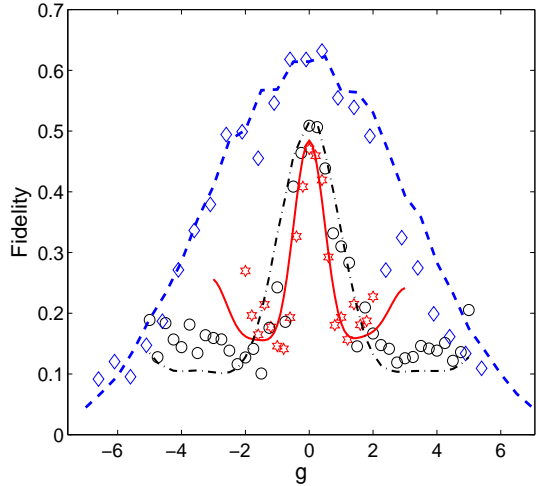}
\caption{(Color Online) Plot showing the fidelity as a function of acceleration. Experimentally measured fidelity  for $\ell=1$ (blue diamonds), $\ell=2$ (black circles) and $\ell=3$ (red stars) due to four kicks of strength $\phi_d\approx 0.6$ followed by a reversal kick of strength $\approx 4\phi_d$.  The lines are the corresponding  fidelity from numerical simulations  with $\Delta\beta=0.06\hbar G $. Note that the horizontal axis is the real acceleration in order to show  the reduction
 in the peak width as $\ell$ increases. }\label{differentl}
\end{figure}

The dynamics of the kicked accelerator  can be described by a Hamiltonian
which in dimensionless units is \cite{fgr}:
\begin{equation}\label{hamiltonian1}
\hat{H}=\frac{\hat{p}^{2}}{2}-\f{\eta}{\tau}\hat{x}
+\phi_{d}\cos(\hat{x})\sum_{q=1}^{\T}\delta (t'-q\tau).
\end{equation}  Here $\hat{p}$ is the momentum in units of  $\hbar G$ (two photon recoils)
that an atom of mass $M$ acquires from $\T$ short, periodic pulses of a standing light wave with a grating vector $G=(4\pi/\lambda)\sin\theta$ ($\theta$ is the angle made by each beam with the vertical). Other variables are the position
$\hat{x}$ (in units of $G^{-1}$), and the continuous time variable $t'$
(integer units). The pulse period $T$ is
scaled by $T_{1/2}=2\pi M/\hbar G^{2}$ (the half-Talbot time) to
give the scaled pulse period $\tau=2\pi T/T_{1/2}$. Here we only consider pulse periods which are integer multiples of $T_{1/2}$, i.e. $\tau=2\pi l$, $\ell$ is integer. The strength of
the kicks is given by $\phi_{d}=\Omega^{2}\Delta t/8\delta_{L}$,
where $\Delta t$ is the pulse length, $\Omega$ is the Rabi
frequency, and $\delta_{L}$ is the detuning of the kicking light
from the atomic transition. Finally the scaled acceleration is defined as $\eta=\f{MgT}{\hbar G}$, with $g$ being the  acceleration of the atoms relative to the standing wave.

In the absence of acceleration, the above Hamiltonian reduces to
the standard kicked rotor system.
Due to  the spatial periodicity of the kicking potential the momentum can be decomposed as
$p=n+\beta$ where $n$ is the integer part of the momentum and
$\beta$ $ (0\leq\beta <1)$ is the quasi-momentum. The
spatial periodicity of the kicking potential only
allows the transition between momenta that differ by an integer
multiple of two photon recoils, $\hbar G$, ensuring the conservation
of quasi-momentum. The dynamics of any single value of the quasi-momentum is the same as that of a rotor known as a $\beta-$rotor.

With a non-zero acceleration, the kicked particle becomes the kicked accelerator and the  quantum dynamics of the system can be understood by applying the one-step operator,
$\hat{\mathcal{U}}_{\beta, \phi_d, \eta}(\T)= e^{-i\phi_d\cos\hat{\theta}} e^{-i\f{\tau}{2}(\hat{\mathcal{N}}+\beta+\eta \T+\eta/2)^2}$, where $\hat{\theta}=\hat{x}\mod(2\pi)$ and
$\hat{\mathcal{N}}=-i\f{d}{d\theta}$ is the angular momentum
operator quantized by integers $n$. $\hat{\mathcal{U}}_{\beta, \phi_d, \eta}(\T)$ is time dependent implying that the   quasi-momentum will no longer be conserved.  However,  its conservation  can be restored by writing
Eq. (\ref{hamiltonian1}) in a freely falling frame using a
gauge transformation. The Hamiltonian then becomes,
\begin{equation}\label{hamiltonian2}
\hat{\mathcal{H}}(\hat{\mathcal{N}}, \hat{\theta},
t')=\f{1}{2}\left(\hat{\mathcal{N}}+\beta+\eta\f{t'}{\tau}\right)^2
+\phi_{d}\cos(\hat{\theta})\sum_{q=1}^{\T}\delta
(t'-q\tau).
\end{equation}

 In the current fidelity experiments, the initial state $|\psi(0)\rangle$ is kicked
$\T$ times,   each  kick having a  strength $\phi_d$. At the end of the $\T^\text {th}$ kick a single pulse  with strength  $\T\phi_d$  is applied. We will refer to this as the ``reversal kick" and it can be implemented by shifting the standing wave by $\lambda_G/2$. Thus the fidelity for a particular $\beta-$rotor is:
  $F(\eta,\T)=|\langle\psi(0)|\hat{\mathcal{U}}_{\beta, \mathrm{t}\phi_d, \eta=0}^{\dagger}\hat{\mathcal{U}}_{\beta, \phi_d, \eta}^{\T}|\psi(0)\rangle|^2$. Following the  technique introduced in \cite{saturation}, the final expression for the fidelity  is then given by,
\begin{align}\nonumber\label{fideq}
F(\eta,\T)&=\bigg|e^{-i\phi(\beta, \eta,
\T)-in_0\ell\pi(2\beta+1)(\T-1)-i\ell\pi n_0\eta
\T^2}\\&J_0\left(\sqrt{(\T\phi_d)^2+\phi_d^2|W_{\T}|^2-2
\mathrm{t}\phi_d^2 \text{Re} W_{\T})}\right)\bigg|^2,
\end{align}
where $p_0=n_0+\beta$ is the initial momentum of the plane wave, $\phi(\beta, \eta, \T)=\ell\pi\sum_{q=0}^{\T-1}
\left(\beta+q\eta +\eta/ 2\right)^2$ is the global phase and
$W_\T(\beta, \eta)=\sum_{q=0}^{\T-1} e^{-i[(2\beta+1)\ell\pi]q-2\ell\pi
i q\eta\T +i\ell\pi\eta q^2}$. In the limit $\eta\rightarrow 0$ for
$\ell=2$ and $\beta=0$, the general result in  Eq. (\ref{fideq}) reduces to the special case considered in \cite{fidelity}. Equation (\ref{fideq}) allows for consideration of situations in which the initial state is a  mixture of
plane waves. Here this state is assumed to have a Gaussian  quasi-momentum distribution with a  $\text{FWHM}= \Delta\beta$. For a given  distribution $\rho (\beta)$ of the quasi-momentum,  the  formula for fidelity is generalized  as:   $F(\eta,\T)=\big|\int_0^1\rho(\beta)\langle\psi(0)|
  \hat{\mathcal{U}}_{\beta, \T\phi_d, \eta=0}^{\dagger}\hat{\mathcal{U}}_{\beta, \phi_d, \eta}^{\T}|\psi(0)
  \rangle d\beta\big|^2$, where the average is computed numerically based on  Eq. (\ref{fideq}) \cite{saturation,nonlinearity}.
    From the global phase term, $\phi(\beta, \eta, \T)$, it can be seen that when $\beta\neq0$ the phase induced by different values of $\eta$  depends not only on the magnitude of $\eta$ but also on its sign.%

Our experiments to investigate this system were performed using a similar set up to that described in \cite{fidelity,shrestha}. A  Bose-Einstein condensate (BEC) of  about 40000 $^{87}$Rb atoms was created
in the $5S_{1/2}$, $ F=1$ level using an all-optical trap technique.
\begin{figure}[htb]
\includegraphics[width=8.8 cm]{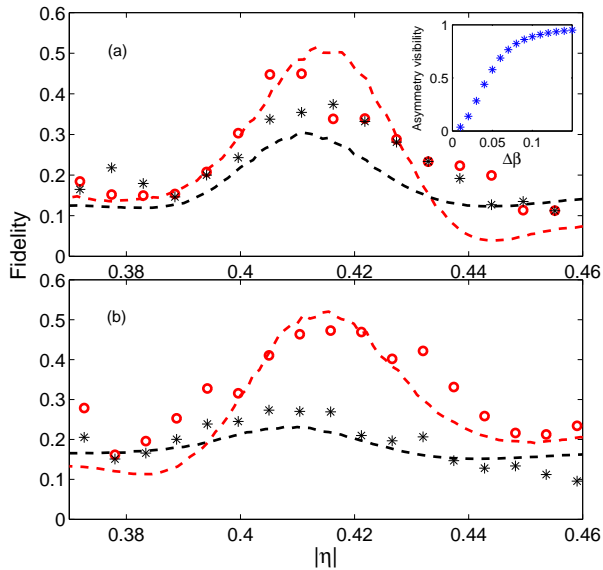}
\caption{(Color Online) Fidelity as a function of $\eta$ for $\tau=4\pi$ and $\beta=0.5$. Red circles and black stars represent experimental fidelity  with negative and positive accelerations respectively. Panels ($a$) and ($b$) correspond to  different $\Delta\beta$ (panel (b) with higher $\Delta\beta$). The measurements were done with four kicks of strength $\phi_d\approx 0.6$ followed by a reversal kick of strength $\approx 4\phi_d$. The dashed lines are the   simulations   for (a) $\Delta\beta = 0.06\hbar G$, and (b) $\Delta\beta = 0.07\hbar G$.  The inset shows the asymmetry visibility (see text)  as a function of $\Delta\beta$. }\label{asymmetry}
\end{figure} Approximately 5 ms
 after  being   released from the trap, the condensate was exposed
 to a  pulsed  horizontal
  standing wave.  This was formed by two laser beams of wavelength
$\lambda=$ 780 nm,  detuned $6.8 $GHz to the red of the atomic transition.
The direction of each beam was aligned at $53^{\text{\textrm{o }}}$
 to the vertical. With these
 parameters the primary QR (half-Talbot time \cite{lepers,talbot}, $\tau=2\pi$) occurred at multiples
 of $51.5 \pm 0.05$ $\mu $s.  Each laser beam   passed through an acousto-optic modulator   driven
  by an arbitrary waveform generator. This enabled  control of the phase,  intensity, and pulse length
   as well as the relative frequency between the kicking beams. Adding two counterpropagating waves differing   in frequency by $\Delta f$ resulted in a standing wave that moved with a velocity $v=2\pi\Delta f/G$.  Since the quasi-momentum $\beta$ of the BEC relative to the standing wave is proportional to $v$,    changing  $\Delta f$  enabled  $\beta$ to be systematically controlled.

The kicking pulse sequence  is  similar to that described in \cite{fidelity}.
The atoms were exposed to a set of $\T$ periodic pulses (forward pulses) each
 of length 1.08 $\mu$s and kicking strength $\phi_d$ followed by the  reversal pulse
 (standing wave displaced by $\lambda_G/2$) with a  strength $\T\phi_d$.
 \begin{figure}[htb]
\includegraphics[width=8.8 cm]{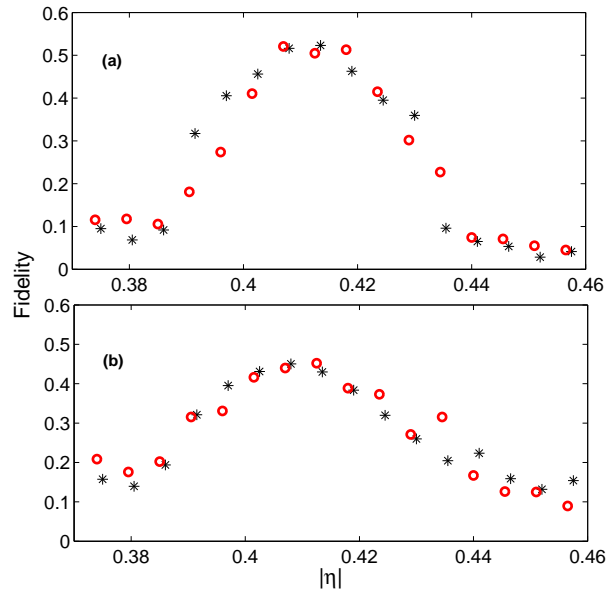}
\caption{(Color Online) Same as in Fig. \ref{asymmetry} but for the center of the quasi-momentum distribution at $\beta=0$. Note that in contrast to Fig. \ref{asymmetry} there is  no  asymmetry between the positive and negative $\eta$'s. }\label{noasymmetry}
\end{figure} We varied the intensity rather
  than the pulse length to change the kicking strength  $\phi_d$.
This was done by
  adjusting the amplitudes of the RF waveforms driving the kicking pulses. This ensured
  that the experiments were always performed  in the Raman-Nath regime
(the distance an atom
   travels during the pulse is much smaller than the spatial period of the potential).
Finally the  kicked atoms were absorption imaged in a time-of-flight experiment and
the fraction of atoms which returned to the initial
momentum state was determined. Experimentally the fidelity was defined as $F=p_0/\sum_n p_n$ where $p_n$ is the number of atoms in the $n^{\text{th}}$ momentum order.  The value of $\Delta\beta$ was varied by changing the  power of the CO$_2$ laser beam which formed the dipole trap used to realize evaporative cooling  in the experiment.  By adjusting the  power of the laser  for the final step in the evaporative sequence we were able to change $\Delta\beta$.

 Figure \ref{longscan} shows the experimentally measured  fidelity   as a function
 of acceleration for $\ell=1$ and initial momentum $\beta=0.5$ due to four kicks each of strength $\phi_d\approx0.6$ followed by a reversal kick of strength  $\phi_d\approx2.4$. Numerical simulations were performed with these experimental parameters under two different conditions. First the black solid line is a simulation in which the reversal pulse is perfect in  amplitude ( amplitude $=\T\phi_d$), and there are no random phase variations in the standing wave that could be caused by vibrations of the optics used to form it. In order to attempt to explain the large deviation of this simulation from the experiment, we also carried out a simulation in which the above experimental imperfections were included (red dashed line). Here we used experimentally realistic values of strength of the reversal kick ($\pm7\% $ from the ideal kick strength) and a random phase variation due to vibrations of $ 0.02\pi$ per pulse. As can be seen the fit to the experiment is quite good, leading us to believe that these effects are the most likely reason for the black curves poor match to the experiment at the $\eta=0$ resonance. In the simulations that follow, we will employ the method used to generate the red dashed curve (with the same parameters for the experimental imperfections).

  Unlike in previous work where only the central resonance was observed \cite{njpfidelity,fidelity}, it is now possible to see that the fidelity has a more complex structure with many resonances away from $\eta=0$.  The validity of the theory for higher resonances at $\ell=2$ and $ 3$ was also tested, the results of which are presented in Fig. \ref{differentl}. Due to  the longer time available for momentum state phases to evolve at the larger $\ell$, the peaks become narrower as $\ell$ is increased. Note that
 the  fidelity  is presented  as a function of real acceleration  in order to show this effect.

We also examined the dependence of the  fidelity  to the sign of $\eta$ (positive and negative acceleration). Asymmetry as predicted by the above theory after Eq. (\ref{fideq}) was observed when the  $\beta-$rotor  distribution was  centered at $\beta=0.5$. It became more prominent as $\Delta\beta$ was increased as shown in Fig. \ref{asymmetry}. Note that  the results correspond to pulse periods, $\tau=4\pi$ ($\ell=2$). The origin of the asymmetry is the different phases $\phi(\beta,\eta,\T)$ induced by the negative and positive values of acceleration. Figure \ref{asymmetry} shows the development of the  asymmetry, both in the experiment and simulations,  as $\Delta\beta$ is increased. The dashed lines are the plot of the simulations with  $\Delta\beta =0.06\hbar G$ and $0.07 \hbar G$ (panels (a) and (b) respectively). 
  An ``asymmetry visibility" defined as $(F(\eta_-)-F(\eta_+))/(F(\eta_-)+F(\eta_+))$ shows an almost linear scaling with the momentum width ($\Delta\beta\leq0.08\hbar G$) of the cloud (see inset). Thus  measurement of the asymmetry may provide a means of determining small $\Delta\beta$ and hence the temperature of ultra-cold atomic clouds.
  Interestingly, the asymmetry goes away if the initial $\beta$ distribution  is chosen  centered at $\beta=0$ as is possible for $\ell=2$  (see  Fig. \ref{noasymmetry}) for the same two $\Delta\beta$'s used in Fig. \ref{asymmetry}. In this case, the distribution is symmetric so that the distribution on the  negative side is identical to that on the positive side. Thus  changing  the sign of  the acceleration, $\eta$, has no effect  on the dynamics. However with the $\beta$ distribution centered at any value other than zero, the distribution is no longer symmetric  and  the effect of $\eta$ will be different for each half of the $\beta$ distribution.

In conclusion, we performed an experimental investigation on the sensitivity of the fidelity to the acceleration by  exposing a BEC to a set of delta-kicked rotor optical pulses followed by a stronger reversal pulse. The experimental results and analytical theory were in good agreement with both showing the presence of multiple fidelity resonances. The width of the central fidelity resonance  was found  to become narrower as the pulse period increased. The importance of the position of the center of the initial momentum distribution  was also explored. When the distribution was centered at some values other than zero, an asymmetry between the fidelity at positive and negative values of acceleration was observed which became more prominent with increasing  $\Delta\beta$.  The asymmetry was optimum for a distribution centered at $\beta=0.5$, disappearing almost completely when the distribution was centered at $\beta=0$. These findings  can be used to determine the  temperature of ultra-cold atoms, based on the scaling of the asymmetry with $\Delta \beta$ (inset in Fig. \ref{asymmetry}).

This work was partially supported by the NSF under
Grant No. PHY-0653494 and by the DFG through FOR760
(WI 3426/3-1), the HGSFP (GSC 129/1), the CQD and the
Enable Fund of Heidelberg University. We thank R. Dubertrand
and I. Talukdar for helpful comments and discussions.

\end{document}